\documentclass[epj,nopacs]{svjour}

\usepackage{amssymb}
\usepackage{epsfig}
\usepackage{eucal}
\usepackage{amsmath}

\newcommand{\taa}[1]{#1\hspace{-.58em}/\hspace{-.09em}} 
\newcommand{\beq}{\begin{equation}} 
\newcommand{\eeq}{\end{equation}}   
\newcommand{\bea}{\begin{eqnarray}}
\newcommand{\eea}{\end{eqnarray}}

\headnote{
\hfill \parbox{4cm}{\tt \normalsize CERN-PH-TH/2004-047 \\ TTP04-05 \\}}

\title{\kern-8pt Nucleon form factors, $\boldsymbol{B}$-meson factories 
and the radiative return
\kern-6pt\thanks{Work supported in part by BMBF under grant number 05HT9VKB0,
EC 5th Framework Programme under contract 
HPRN-CT-2002-00311 (EURIDICE network),
Polish State Committee for Scientific Research (KBN)
under contract 2 P03B 017 24,  BFM2002-00568,
Generalitat Valenciana under grant GRUPOS03/013,
and MCyT under grant FPA-2001-3031.}}

\author{Henryk Czy\.z\inst{1}
\thanks{\email{czyz@us.edu.pl}} 
\and
Johann H. K\"uhn\inst{2}
\thanks{\email{johann.kuehn@physik.uni-karlsruhe.de}}
\and
El\.zbieta Nowak\inst{1}
\thanks{\email{ela@higgs.phys.us.edu.pl}}
\and
Germ\'an Rodrigo\inst{3,4}
\thanks{\email{german.rodrigo@cern.ch}}
}
\institute{
Institute of Physics, University of Silesia,
PL-40007 Katowice, Poland. \and
Institut f\"ur Theoretische Teilchenphysik,
Universit\"at Karlsruhe, D-76128 Karlsruhe, Germany.
\and
Department of Physics, CERN, Theory Division, CH-1211 Geneva 23, Switzerland 
\and
Instituto de F\'{\i}sica Corpuscular, E-46071 Valencia, Spain.
}

\date{Received: March 5, 2004}

\begin{document}

\abstract{The feasibility of a  measurement of the electric and magnetic
nucleon form  factors at $B$-meson factories through  the radiative return
is studied.   Angular distributions allow a separation of the contributions
from  the two  form factors.  The  distributions are  presented for  the
laboratory and the hadronic rest  frame, and the advantages of different
coordinate  systems  are investigated.  It  is  demonstrated that  $Q^2$
values up  to 8 or  even 9~GeV${}^2$ are  within reach. The  Monte Carlo
event generator PHOKHARA is extended to nucleon final states, and results
are presented which  include Next-to-Leading Order radiative corrections
from initial-state radiation. The impact  of angular cuts  on rates and
distributions is  investigated and the relative  importance of radiative
corrections is analysed.
}

\def\Li{\hbox{Li}}                                            
 

\maketitle

\section{Introduction}
The importance of measuring the nucleon form factor has been
repeatedly emphasized in the literature (e.g. 
\cite{iachello,brodsky,Baldini}  and refs therein).
Recent experiments at Jefferson Lab have explored the ratio of electric and
magnetic form factors in the space-like region \cite{JLab1,JLab2}
up to ~5.6~GeV$^2$, using the recoil polarization method; their results show 
disagreement with data obtained with the Rosenbluth method 
(For an extensive review see~\cite{arrington}. For recent suggestions of a 
possible explanation of this discrepancy through corrections from two-photon
exchange see e.g.~\cite{Guichon:2003qm,Chen:2004tw}.).
Data in the time-like region based on electron--positron annihilation
into proton--antiproton (and neutron--antineu\-tron) final states, and the
inverse reactions, only extend to 6~GeV${}^2$ and 14~GeV${}^2$ respectively. 
These latter measurements exhibit fairly large errors and do not
provide a separation of the contributions from the two form factors.
To explore the validity of the different model predictions, improved
measurements over a wide kinematic range are desirable.

In the present paper the potential of the radiative return at $B$-meson
factories will be explored. As shown below, these measurements will
cover the region from threshold up to 8 or even 9 GeV${}^2$ with
sufficient counting rates. This region is also accessible at the
Beijing storage ring and is complementary to the one at Cornell with
energies above the $J/\psi$ resonance.

Not surprisingly, many aspects of the radiative return are quite similar
to those of the reaction $e^+e^-\to p\bar p$. In particular it is
again the square of the electric and magnetic form factors, which can
be determined separately through the analysis of angular distributions.

Radiative corrections are indispensable for a precise interpretation
of the data. Furthermore, given the limited acceptance and the
asymmetric kinematic configuration at present $B$-meson factories, 
a Monte Carlo event generator is required to demonstrate the feasibility 
of the measurement. The present analysis is based on an extension of the
generator PHOKHARA 3.0 \cite{Czyz:PH03}, which was originally constructed 
to simulate two-pion and two-muon \cite{Rodrigo:2001kf},
and later four-pion \cite{Czyz:2002np,Czyz:2000wh}, production through 
the radiative return \cite{Binner:1999bt,Zerwas}
and which includes next-to-leading order (NLO)
radiative corrections \cite{Rodrigo:2001jr,Kuhn:2002xg}.
The new version of PHOKHARA (PHOKHARA 4.0) will include, besides of the 
nu\-cle\-on--antinucleon final states, final-state radiative corrections
(FSR) to $\mu^+\mu^-$ production at NLO \cite{in_preparation}.

For the present case we consider photon emission from the initial
state only (ISR). As discussed below in more detail, final-state radiation
is completely negligible if the reaction is considered at a 
centre of mass (cms) collider energy of 10.5~GeV.

The choice of form factors made by the program is directly taken 
from \cite{iachello,ia-ja-la}. It is, however, set up
in a modular form such that the two form factors can be easily
modified and replaced by the user, if required.

The program also allows a more detailed study of angular distributions, 
including the effects of cuts on photon and proton acceptance.
Indeed we will find that these distributions are quite sensitive to the
ratio of the form factors, and the extraction of this ratio seems to
be feasible over a fairly large kinematic range.

\section{Electric and magnetic form factors and the
radiative return}\label{meth}

The matrix element of the electromagnetic current governing
nucleon--antinucleon production is conventiona\-lly written in
terms of the Dirac ($F_1^N$) and Pauli ($F_2^N$) form factors  
\begin{equation}
 J_\mu =  - i e \cdot \bar u(q_2)\left(F_1^N(Q^2)\gamma_\mu
 - \frac{F_2^N(Q^2)}{4 m_N} \left[\gamma_\mu,\taa Q \ \right]
 \right) v(q_1)~,
\end{equation}
where $N$ stands for proton or neutron. The nucleon and antinucleon
momenta are denoted by $q_1$ and $q_2$ respecti\-vely, and 
$Q=q_1+q_2$.

Cross sections and distributions for $e^+e^-\to N\bar N$
are conveniently expressed in terms of the
electric and magnetic Sachs form factors \cite{Sachs}:
\begin{equation}
G_M^N = F_1^N + F_2^N~, \qquad  G_E^N = F_1^N + \tau F_2^N~,
\end{equation}
with $\tau =Q^2/4m^2_N$,
which will also lead to compact formulae for the radiative return.
Both proton and neutron form factors can be decomposed into
isoscalar  and isovector contributions (see e.g. \cite{ia-ja-la}):
\begin{align}
F^p_{1,2}= F_{1,2}^s + F_{1,2}^v~, \qquad 
F^n_{1,2} = F_{1,2}^s - F_{1,2}^v~.
\end{align}

Let us start with a qualitative discussion of form factor measurements
through the radiative return, based on the leading order process
\begin{equation}
e^+(p_1)+e^-(p_2) \to \bar{N}(q_1)+ N(q_2)+\gamma(k)~.
\label{proc}
\end{equation}
From the simple analytical results it will be straightforward to
evaluate production rates, to understand the qualitative aspects of
angular distributions, and to develop strategies for the separation of
the electric and the magnetic form factor.

The differential cross section for  
reaction (\ref{proc}) (with ISR only) can be written as
\begin{equation}
d\sigma = \frac{1}{2s}L_{\mu\nu}H^{\mu\nu}d\Phi_2(p_1+p_2;Q,k)
d\Phi_2(Q;q_1,q_2)\frac{dQ^2}{2\pi}~,
\end{equation}
where $L_{\mu\nu}$ and $H^{\mu\nu}$ are the leptonic and hadronic tensors 
respectively. For notation, definitions and an explicit form of the 
leptonic tensor, see for instance \cite{Czyz:2000wh,Kuhn:2002xg}. 
The hadronic tensor is given by
\bea
H_{\mu\nu}&=&2|G_M^N|^2(Q_\mu Q_\nu - g_{\mu\nu} Q^2) \nonumber \\ 
 &&- \frac{8\tau}{\tau -1}\left(|G_M^N|^2-\frac{1}{\tau}|G_E^N|^2\right)q_\mu q_\nu~,
\label{had_tens} 
\eea
where $q=(q_2-q_1)/2$.
From the explicit form of the hadronic tensor it becomes apparent that
the relative phase between $G_E^N$ and $G_M^N$ cannot be measured in
this experiment (i.e. as long as the spin of the nucleon remains
unmeasured). This is independent of the detailed form of the leptonic tensor.
In particular phases from higher order ISR or (longitudinal or
transversal) beam polarization affect only the leptonic tensor
and thus do not alter this conclusion.

Integrating over the whole range of nucleon angles and the
photon azimuthal angle, and restricting the photon polar angle
within $\frac{m_e}{\sqrt{s}} \ll
\theta_\gamma^{\rm min} < \theta_\gamma < \pi -\theta_\gamma^{\rm min} $,
the differential cross section factorizes into the cross section for
electron--positron annihilation into hadrons and a flux factor that
depends on $s$ and $Q^2$ only \cite{Binner:1999bt}:
\begin{equation}
  \frac {{\rm d}\sigma}{{\rm d} Q^2} = \frac
{4 \alpha^3}{3 s Q^2 } R(Q^2) 
 \biggl\{ \frac {(1+\frac{Q^4}{s^2})}{(1-\frac{Q^2}{s})}  
\log \frac {1+c_m} 
{1-c_m}  
- \left(1-\frac{Q^2}{s}\right) c_m \biggl\}~,
\label{rad_ret}
\end{equation}
where $c_m=\cos\theta_\gamma^{\rm min}$, $m_e$ denotes the electron mass,
\begin{equation}
R(Q^2)=\frac{\sigma(e^+e^-\rightarrow N \bar{N})}{\sigma_{\rm point}}=
\frac{\beta_N}{2}\left(2|G_M^N|^2+\frac{1}{\tau}|G_E^N|^2\right)~,
\label{rr}
\end{equation}
with $\sigma_{\rm point}$ the lowest order muonic cross section
and  $\beta_N^2 = 1 - 4m_N^2/Q^2$.

After integration over the whole range of nucleon angles,
the  separation of electric and magnetic form  factors is no longer
feasible. However, the modulus of their ratio can be determined 
from the study of properly chosen angular distributions. 
The fully differential cross section is essentially given by 
\begin{align}
&  L_{\mu\nu}H^{\mu\nu} = \frac{(4\pi\alpha)^3}{Q^2} \biggl\{
\biggr(|G_M^N|^2-\frac{1}{\tau}|G_E^N|^2 \biggr)\label{eq:LH}
  \\&
\kern +20 pt \times \frac{32 s}{\beta_N^2(s-Q^2)} 
\biggr(\frac{1}{y_1}+\frac{1}{y_2}\biggr)  
\biggr(\frac{(p_1\cdot q)^2 +(p_2\cdot q)^2}{s^2} \biggr)  \nonumber \\&
 + 2\biggr(|G_M^N|^2+\frac{1}{\tau}|G_E^N|^2 \biggr) 
\biggr[ \biggr(\frac{1}{y_1}+\frac{1}{y_2}\biggr) \
\frac{(s^2+Q^4)}{s(s-Q^2)} - 2  \biggr]
\biggl\}~,
\nonumber
\end{align}
where $y_{1,2}=\frac{s-Q^2}{2s}(1\mp \cos\theta_{\gamma})$.

The separation of the form factors will therefore have to rely on the
peculiar dependence of the differential cross section on the nucleon and
antinucleon momenta separately.

The close connection between this analysis and the one based on the
reaction $e^+e^-\to N\bar N$ becomes even more apparent once the result
is expressed in terms of the polar and azimuthal angles $\hat\theta$ and
$\hat\varphi$, which characterize the nucleon direction in the rest frame
of the hadronic system, with the $z$-axis opposite to the direction of
the real photon momentum and the $y$-axis in the plane spanned by the
beam and the photon direction (see Appendix A):
\begin{align}
&  L_{\mu\nu}H^{\mu\nu} =    \label{eq:LH2} 
        \\& 
= \frac{(4\pi\alpha)^3}{Q^2} \biggl\{
\biggr(|G_M^N|^2-\frac{1}{\tau}|G_E^N|^2 \biggr)  
\frac{4 Q^2}{(s-Q^2)}
\biggr(\frac{1}{y_1}+\frac{1}{y_2}\biggr)\nonumber \\& 
\phantom{00} \times \biggr( (\beta\gamma\cos\hat\theta)^2 + 
   (\gamma\cos\theta_\gamma\cos\hat\theta -
       \sin\theta_\gamma\sin\hat\theta\sin\hat\varphi)^2\Biggl)
\nonumber \\&
 + 2\biggr(|G_M^N|^2+\frac{1}{\tau}|G_E^N|^2 \biggr) 
\biggr[ \biggr(\frac{1}{y_1}+\frac{1}{y_2}\biggr) \
\frac{(s^2+Q^4)}{s(s-Q^2)} - 2  \biggr]
\biggl\}~, \nonumber
\end{align}
where $\gamma=(s+Q^2)/2\sqrt{s Q^2}$ and $\beta=(s-Q^2)/(s+Q^2)$
characterize the boost from the laboratory to the hadronic rest frame.
In the limit $Q^2\ll s$, this can be approximated by
\begin{align}
L_{\mu\nu}H^{\mu\nu} =& \frac{(4\pi\alpha)^3}{Q^2} 
\frac{(1+\cos^2\theta_\gamma)}{(1-\cos^2\theta_\gamma)}
\label{eq:LH3}
\\& 
\times 
4\left(|G_M^N|^2(1+\cos^2\hat\theta) + \frac{1}{\tau}|G_E^N|^2\sin^2\hat\theta\right)~.
\nonumber
\end{align}

An alternative choice of the coordinate system, which is obtained
through an additional rotation around the $x$-axis in the hadronic rest
frame, leads to a diagonal form of the leptonic tensor. 
In this frame the angular distribution of the baryons simplifies even further.
This formulation is discussed in detail in the appendix. For small  
$Q^2/s$ the two frames nearly coincide.

It is instructive to compare the angular distribution in Eq.~(\ref{eq:LH3}) 
with the angular distribution from  $e^+e^-\to N\bar N$:
\begin{equation}
\frac{d\sigma}{d\Omega}=\frac{\alpha^2 \beta_N}{4 Q^2}
\left(|G_M^N|^2(1+\cos^2\theta) + \frac{1}{\tau}|G_E^N|^2\sin^2\theta\right)~.
\label{sigdir}
\end{equation}
The close relation between Eqs.~(\ref{eq:LH3}) and (\ref{sigdir}) 
is clearly visible.

\hskip-0.2cm 
As is already evident from the hadronic tensor in Eq.~(\ref{had_tens}),
the phases of the form factors are, however, not accessible by this method.
Their determination would require the measurement of the polarization of the
nucleons in the final state \cite{DDR,Rock}. PHOKHARA can easily be
extended to describe this additional degree of freedom.
This option is of particular interest for $\Lambda \bar \Lambda$
production, with its self-analysing decay mode.

As emphasized above, only ISR is included in the pre\-sent version of the
program. FSR with one photon only emitted from the hadronic system is 
described by the amplitude for $e^+e^-\to\gamma^*\to N\bar N \gamma$, 
with an invariant mass of the virtual photon of 10.5 GeV. This
option is, however, strongly suppressed by the proton form factor. In
fact, the arguments are quite similar to those for pion pair production,
where FSR in leading order and at 10.5 GeV is completely negligible.

As will be discussed below, the event rate drops rapidly with increasing
$Q^2$ and only few events will be observed for invariant hadron masses
around 3~GeV. However, the rate for baryon pair production from $J/\psi$
decays will be significantly enhanced, all considerations given above
for continuum production do apply, and a precise measurement of the
branching ratio of $J/\psi$ into baryon pairs and the relative strength
of magnetic and electric coupling is within reach.

\section{Nucleon form factors and their implementation in PHOKHARA}

The parametrization of the form factors used in this paper and listed 
below in detail is adopted from Ref. \cite{ia-ja-la},
with the analytical continuation to the time-like region taken from 
 \cite{iachello} and \cite{brodsky}. The isoscalar and isovector
components of Dirac and Pauli form factors are given by:
\begin{equation}
F_1^s = \frac{g(Q^2)}{2}\biggr[ (1-\beta_{\omega}-\beta_{\phi}) 
-\beta_{\omega}\cdot T_\omega
-\beta_{\phi}\cdot  T_\phi\biggr]~,
\label{Fs}
\end{equation}
\begin{equation}
F_2^s = \frac{g(Q^2)}{2}\biggr[ (0.120+\alpha_{\phi}) 
 \cdot T_\omega
-\alpha_{\phi}\cdot T_\phi \biggr]~,
\end{equation}
\begin{equation}
F_1^v = \frac{g(Q^2)}{2}\biggr[ (1-\beta_{\rho}) 
-\beta_{\rho} \cdot T_\rho \biggr]~,
\end{equation}
\begin{equation}
F_2^v = \frac{g(Q^2)}{2}\biggr[ -3.706 \cdot T_\rho \biggr]~,
\label{Fv}
\end{equation}
with 
\begin{equation}
T_\rho = \frac{m^2_{\rho} + 8\Gamma_{\rho}m_{\pi}/\pi}
{Q^2-m^2_{\rho}+(Q^2-4m^2_{\pi})\Gamma_{\rho}\alpha(Q^2)/m_{\pi}} \ ,
\end{equation}
\begin{equation}
\alpha(Q^2) = \biggr( 1 - x^2  \biggr)^{1/2}
\biggl\{ \frac{2}{\pi}\log \biggr( \frac{1+\sqrt{1-x^2} }
{x} \biggr) - i \biggl\} \ ,
\end{equation}
and
\begin{equation}
T_{\omega,\phi} = \frac{m^2_{\omega,\phi}}{Q^2-m^2_{\omega,\phi}}~, \ 
g(Q^2) = \frac{1}{(1-\gamma e^{i\theta}Q^2)^2}~, \ 
x = \frac{2m_{\pi}}{\sqrt{Q^2}}~.
\label{gg}
\end{equation}

\begin{figure}
\begin{center}
\hskip -1cm \epsfig{file=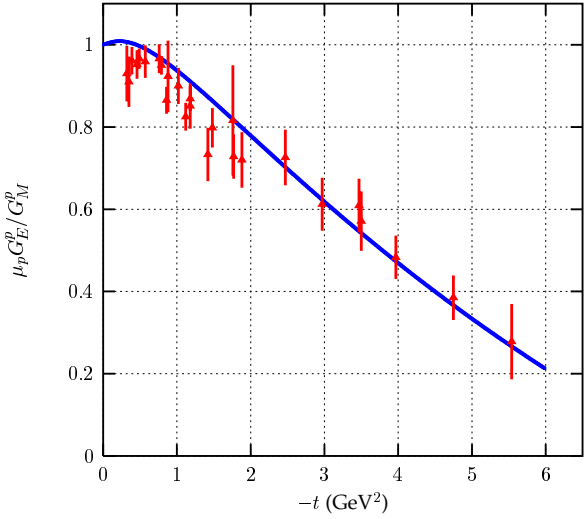,width=8cm} 
\end{center}
\caption{The ratio of proton form factors in the
space-like region as predicted
by the model \cite{ia-ja-la} (solid line) and data from
JLab  \cite{JLab1,JLab2}. $\mu_p$ is the proton magnetic moment. }
\label{fig:13}
\end{figure}

\begin{figure}
\begin{center}
\hskip -1cm \epsfig{file=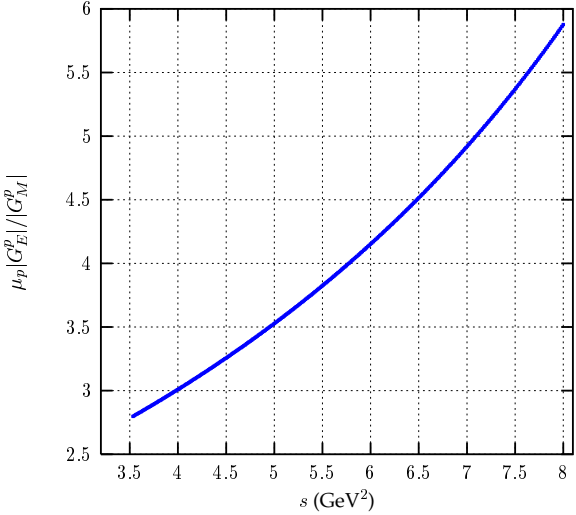,width=8cm} 
\end{center}
\caption{Modulus of the ratio of proton form factors in the
time-like region as predicted by the model \cite{ia-ja-la}.}
\label{fig:14}
\end{figure}

The values of the parameters (dimensionful quantities in units of GeV) are 
$\beta_{\rho}=0.672$, $\beta_{\omega}=1.102$, $\beta_{\phi}=0.112$,
$m_{\phi}=1.019$, $m_{\rho}=0.765$, $m_{\omega}=0.784$, 
$\alpha_{\phi}=-0.052$, $\Gamma_{\rho}=0.112$, $\gamma = 0.25$.
The angle $\theta$ in Eq.~(\ref{gg}) is set to
$\theta =\pi/4$ or $ \theta = 19\pi/64$  as recommended in \cite{iachello}.
More recent fits to the data
in the space-like region are listed in \cite{gari_kru,lomon}.
However, in view of the large uncertainty of the data the discrimination
between the different parametrizations is not yet possible.
It should be emphasized that the original
fit of \cite{ia-ja-la} agrees well with the ratio of the form factors
measured at Jefferson Lab \cite{JLab1,JLab2}, as shown in Fig.~\ref{fig:13}. 
In the time-like region, the ratio of the form factors is predicted to
increase with $Q^2$ (Fig.~\ref{fig:14}), in contrast to the behaviour
in the space-like region.

\begin{figure}
\begin{center}
\hskip -1cm \epsfig{file=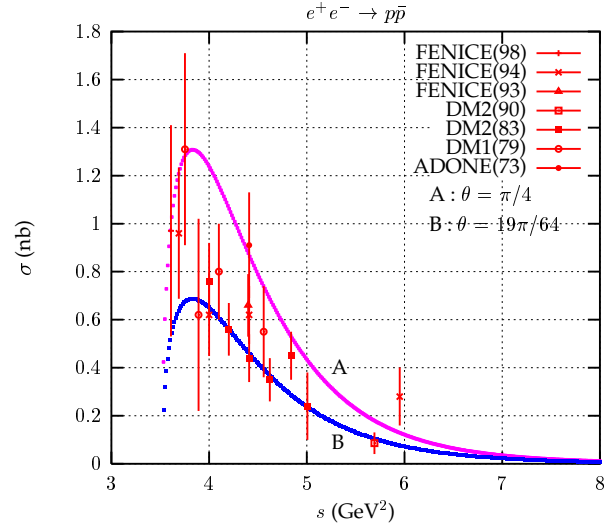,width=8cm} 
\end{center}
\caption{Comparison of the measured 
\cite{antonelli98,antonelli94,antonelli93,bisello90,bisello83,delcourt79,castellano73}
$e^+e^- \to p\bar p$ cross section with the model from 
Ref.~\cite{ia-ja-la}. Predictions are given
for two different values ($\pi/4$ - curve A  and $19\pi/64$ - curve B)
of the parameter $\theta$ (see Eq.~(\ref{gg})).}
\label{fig:1}
\end{figure}

\begin{figure}
\begin{center}
\vskip-0.3cm
\vskip+0.5cm
\hskip-0.5cm \epsfig{file=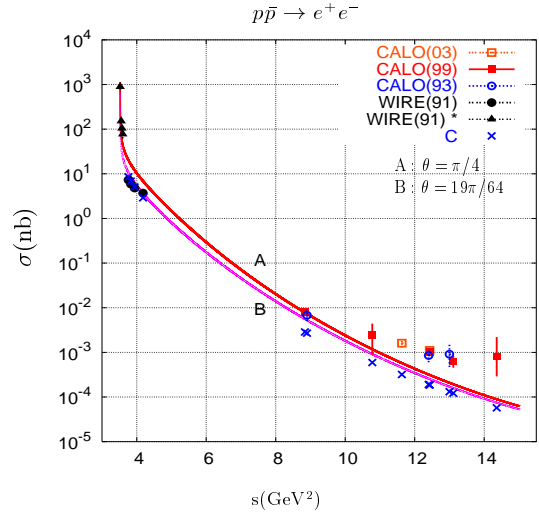,width=7.4cm} 
\end{center}
\caption{Comparison of the measured 
\cite{andreotti03,ambrogiani99,armstrong93,bardin91,bardin91a}
$ p\bar p \to e^+e^-$ cross section with the model 
from Ref.~\cite{ia-ja-la}. Predictions are given
for two different values ($\pi/4$ - curve A  and $19\pi/64$ - curve B)
of the parameter $\theta$ (see Eq.~(\ref{gg})). The curves were obtained
without imposing any angular cuts. Theoretical results with cuts appropriate
for each experiment are shown as crosses. }
\label{fig:2}
\end{figure}

\begin{figure}
\begin{center}
\hskip -1cm \epsfig{file=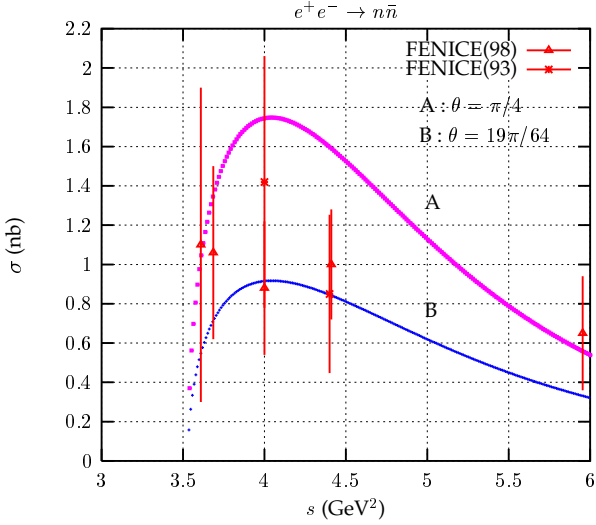,width=8cm} 
\end{center}
\caption{Comparison of the measured 
\cite{antonelli98,antonelli93}
$e^+e^- \to n\bar n$ cross section with the model 
from Ref.~\cite{ia-ja-la}. Predictions are given
for two different values ($\pi/4$ - curve A  and $19\pi/64$ - curve B)
of the parameter $\theta$ (see Eq.~(\ref{gg})).}
\label{fig:3}
\end{figure}

Data giving information about the form factors in the time-like region
originate from the reactions
$e^+e^- \to p\bar p$, $p\bar p \to e^+e^-$  and $e^+e^- \to n\bar n$ 
and are shown in  Figs.~\ref{fig:1}--\ref{fig:3}. 
Only measurements of the total cross section are available.
Form factors, if available at all, are extracted  under the assumption 
$G_M^p = \mu_p G_E^p$, and no true separation of the
electric and magnetic form factors has been performed, a consequence of
the low event rates. Within the large experimental uncertainties the data 
are in reasonable agreement with the model predictions.
The poor experimental knowledge can be substantially improved by using
the radiative return \cite{Binner:1999bt,Zerwas} at  $B$-meson
factories. Even the separation of electric and magnetic form factors
may be envisaged in order to resolve the present discrepancies
\cite{arrington} in the measurement of the proton form factors. 

The form factors defined in Eqs.~(\ref{Fs})--(\ref{Fv}) were 
im\-ple\-men\-ted in the
event generator  PHOKHARA \cite{Czyz:2002np,Czyz:PH03,Rodrigo:2001kf},
providing a powerful tool for the experimental and theoretical ana\-lysis
of the processes $e^+e^-\to p\bar p \gamma$ and $n \bar n \gamma$ at NLO.
Real and virtual FSR contributions are not
(yet) implemented. They might, however, become
important in conjunction with hard ISR (lowering the effective $Q^2$) for
$Q^2$ close to the  $p\bar p$ threshold, where the final-state 
Coulomb interaction is sizeable. 

The technical correctness of the implementation
was checked first by confronting the $Q^2$ distribution of events
generated by PHOKHARA 4.0 at LO with the corresponding 
analytical prediction in Eq.~(\ref{rad_ret}).
The test was successfully
performed with an agreement better than 0.1\% for $Q^2 < 8$ GeV$^2$ (the
region were the cross section is well accessible to experiment). 
At NLO the standard tests for the independence of the cross section on the
`soft--hard' separation parameter were performed again with a precision
better than 0.1\%. They guarantee that the implementation of the form 
factors in the one-photon and two-photon parts of the generator is 
consistent and, together with the LO test, support the correctness of
the whole implementation.

\section{Form factor measurements at $\boldsymbol{B}$-meson factories}

\begin{figure}
\begin{center}
\hskip -1cm \epsfig{file=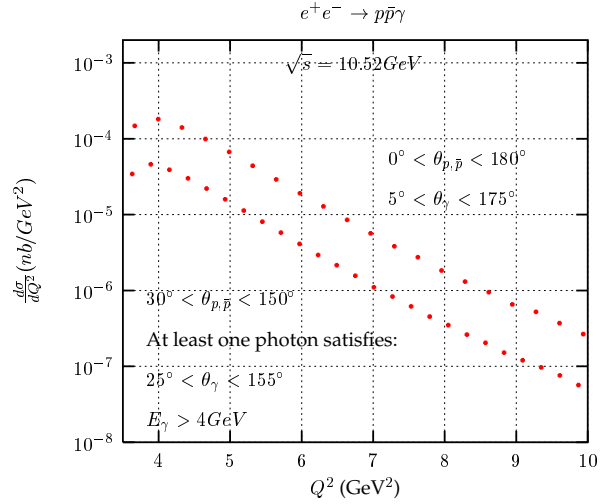,width=8cm} 
\end{center}
\caption{Differential cross section of the process 
$e^+e^- \to p \bar p \gamma (\gamma)$ for two different sets of cuts.}
\label{fig:5}
\end{figure}

\begin{figure}
\begin{center}
\hskip -1cm \epsfig{file=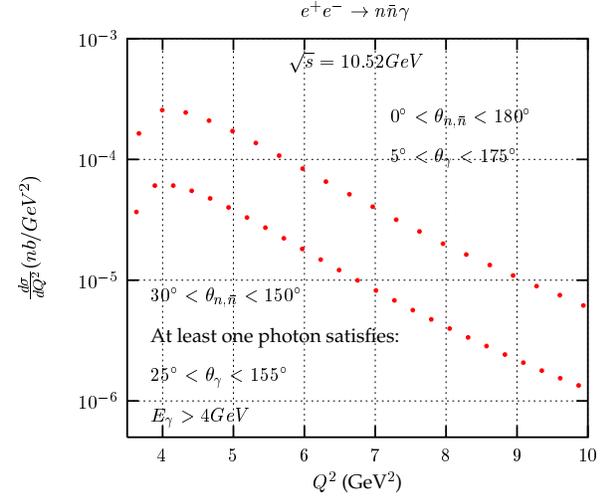,width=8cm} 
\end{center}
\caption{Differential cross section of the process 
$e^+e^- \to n \bar n \gamma (\gamma)$ for two different sets of cuts.}
\label{fig:6}
\end{figure}

The differential cross section as a function of $Q^2$, as observed through
the radiative return, is shown in  Figs.~\ref{fig:5} and \ref{fig:6}.
The upper curve represents the cross section after integration over all
proton angles and over photon angles down to $5^\circ$ w.r.t. the beam
direction. The lower curves are valid for cuts corresponding
to the BaBar acceptance region transformed to the $e^+e^-$ cms. 
The cross sections corresponding to the lower curves, integrated
over $Q^2<$ 10 GeV$^2$, amount to 59.3 fb for protons and to 
125 fb for neutrons.
With a luminosity of over 130 fb$^{-1}$ accumulated at $B$-factories
already now, the event rate may be large enough to extract and separate
$|G_M^N|$ and $|G_E^N|$ as proposed above.  

\begin{figure}
\begin{center}
\epsfig{file=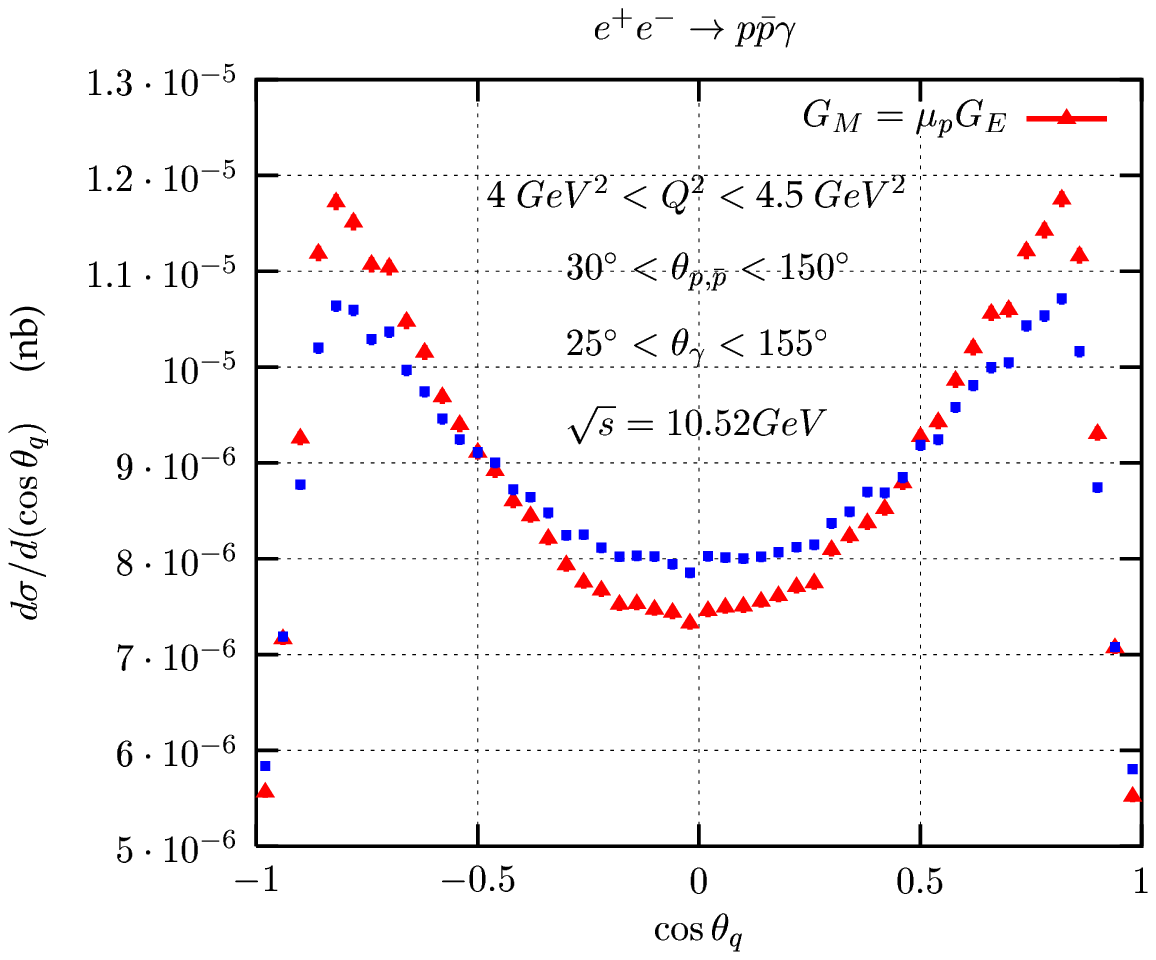,width=8.2cm} 
\epsfig{file=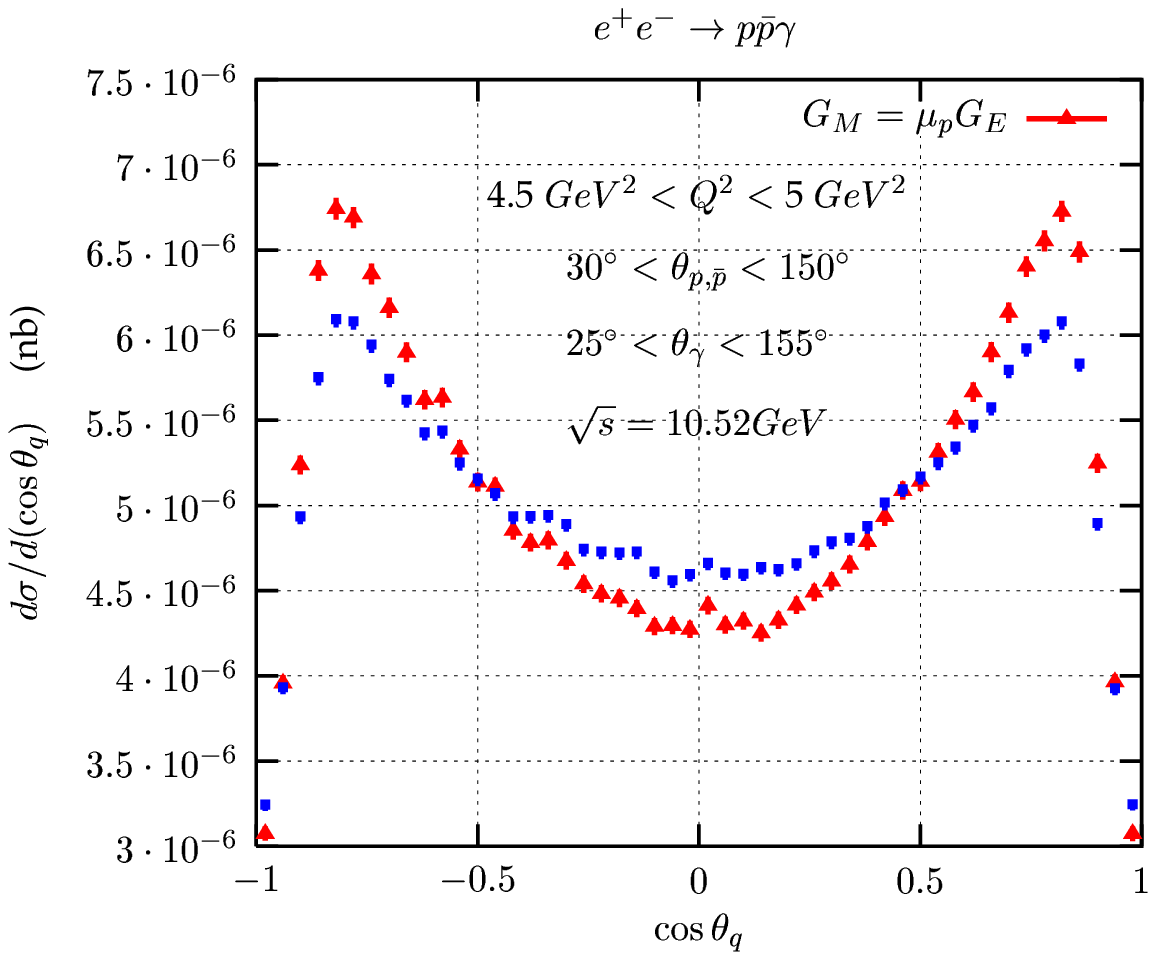,width=8.2cm} 
\epsfig{file=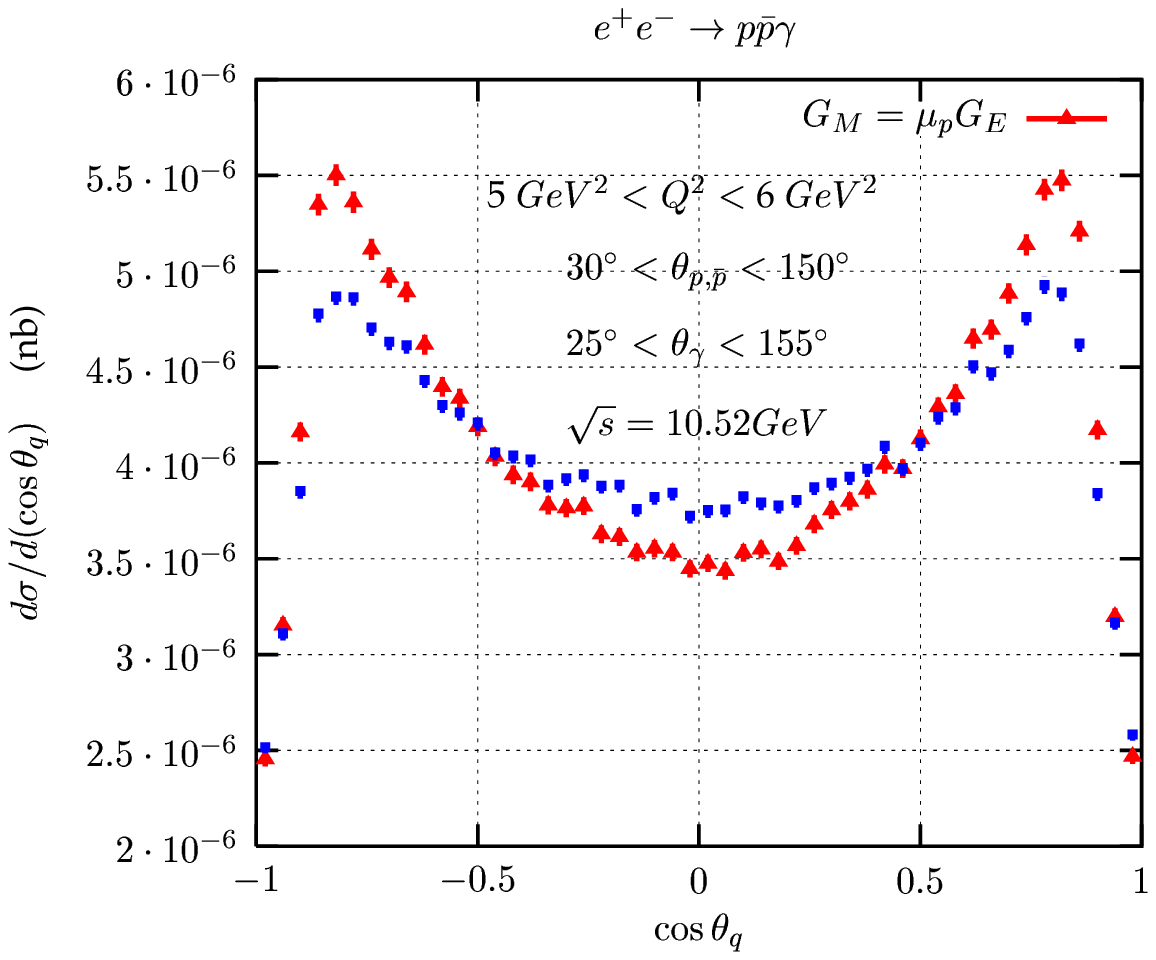,width=8.2cm} 
\end{center}
\caption{Angular distributions in the polar angle of vector
${\bf q}$ for three different ranges of $Q^2$, with and without the constraint
$G_M^p = \mu_p G_E^p $ (see text for a more detailed explanation).}
\label{fig:15}
\end{figure}

In Fig.~\ref{fig:15} we show the angular
distributions in $\theta_q$ (the polar angle of the vector ${\bf q}$)
for different ranges of $Q^2$. To explore the sensitivity to the form 
factors individually we compare the differential cross sections obtained
for the model described above with the distribution obtained under
the assumption $G_M^p = \mu_p G_E^p $, subject, however, to
the constraint that  $\sigma(e^+e^-\rightarrow p \bar{p})$
(see Eq.~(\ref{rr})) remains unaffected. Without that constraint the predicted
cross section would be up to ten times larger, in disagreement with
existing data. 
The distributions are markedly different and the discrimination between
different assumptions about the ratio $G_M^p /G_E^p $
seems feasible.

\begin{figure}
\begin{center}
\epsfig{file=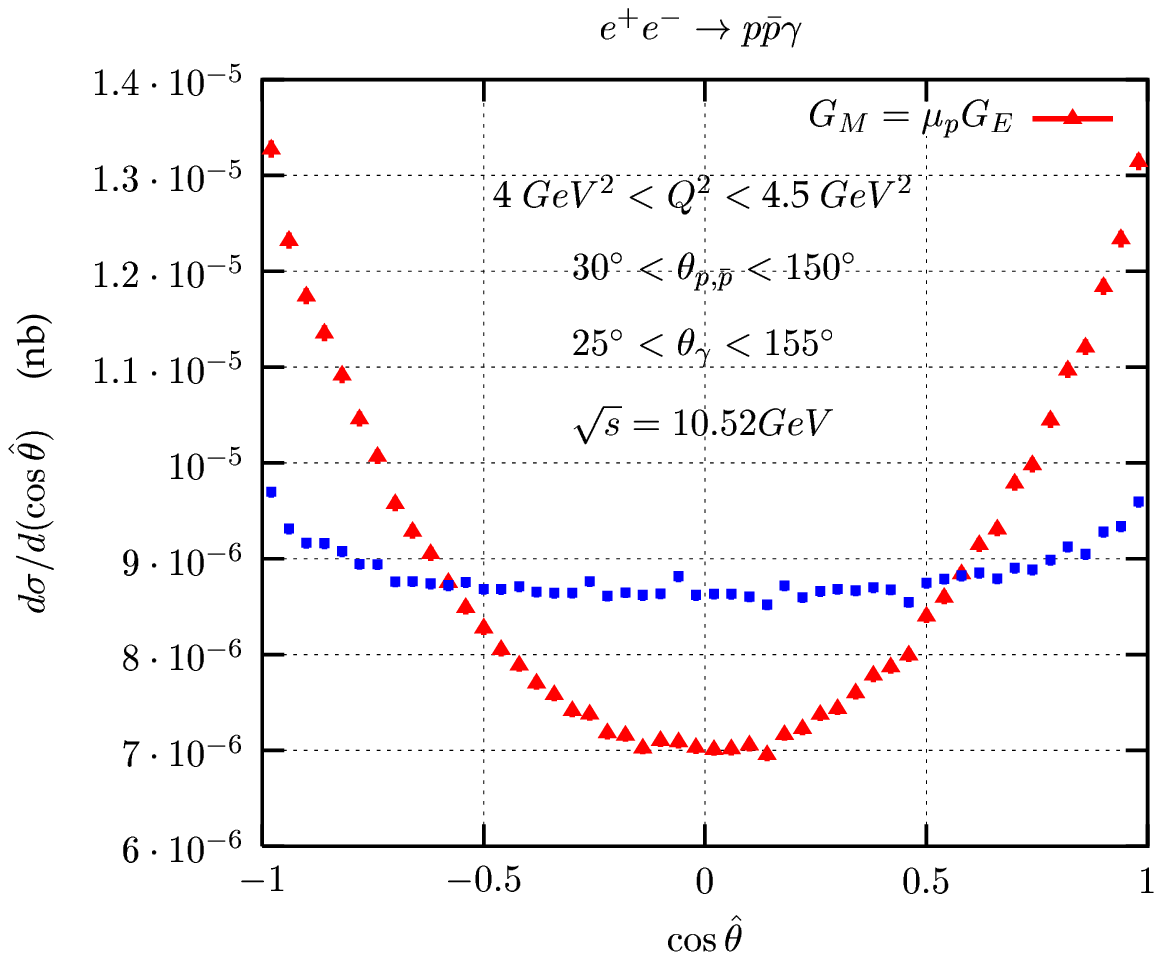,width=8.2cm} 
\epsfig{file=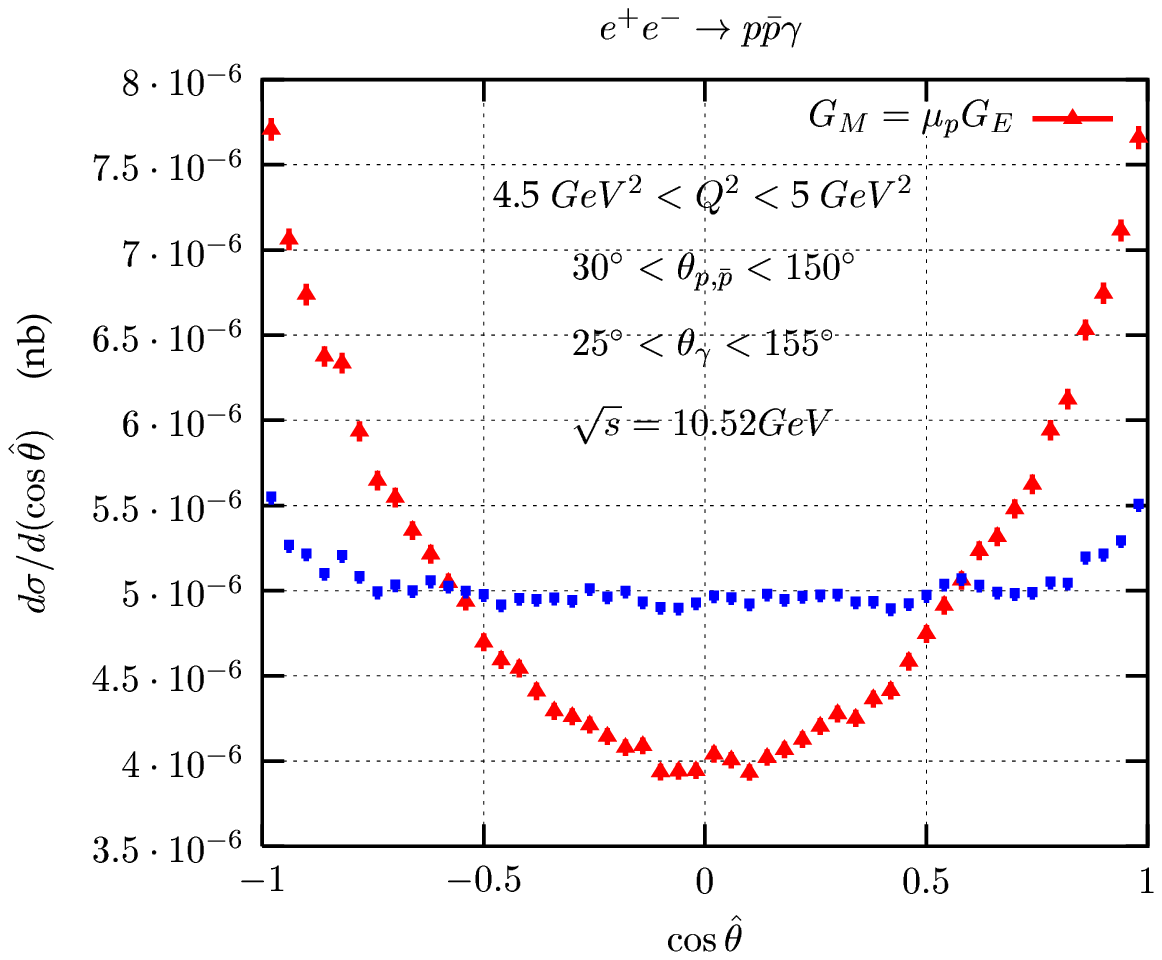,width=8.2cm} 
\epsfig{file=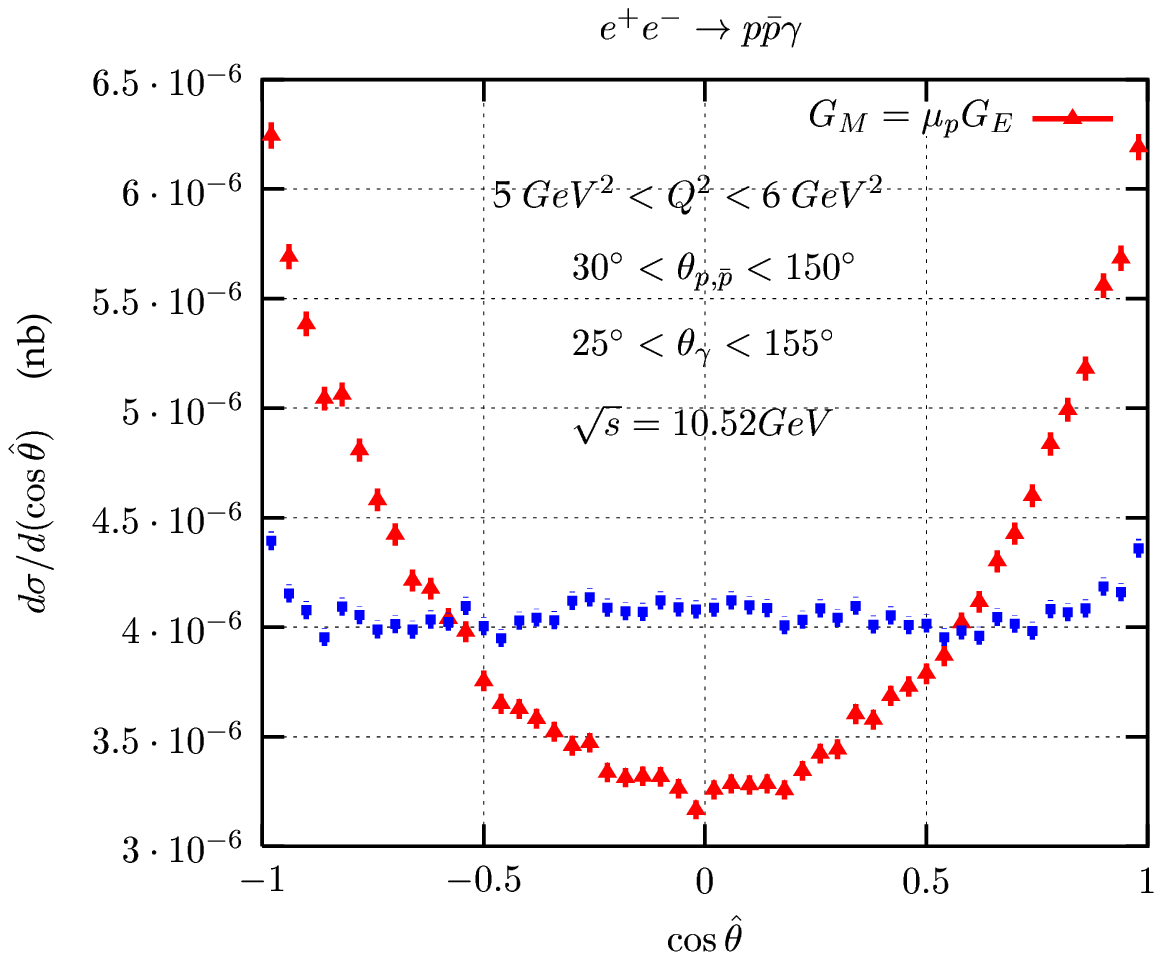,width=8.2cm} 
\end{center}
\caption{Angular distributions in the polar angle of the proton
for three different ranges of $Q^2$, with and without the constraint
$G_M^p = \mu_p G_E^p $ (see text for a more detailed explanation).}
\label{fig:15a}
\end{figure}

\begin{figure}
\begin{center}
\epsfig{file=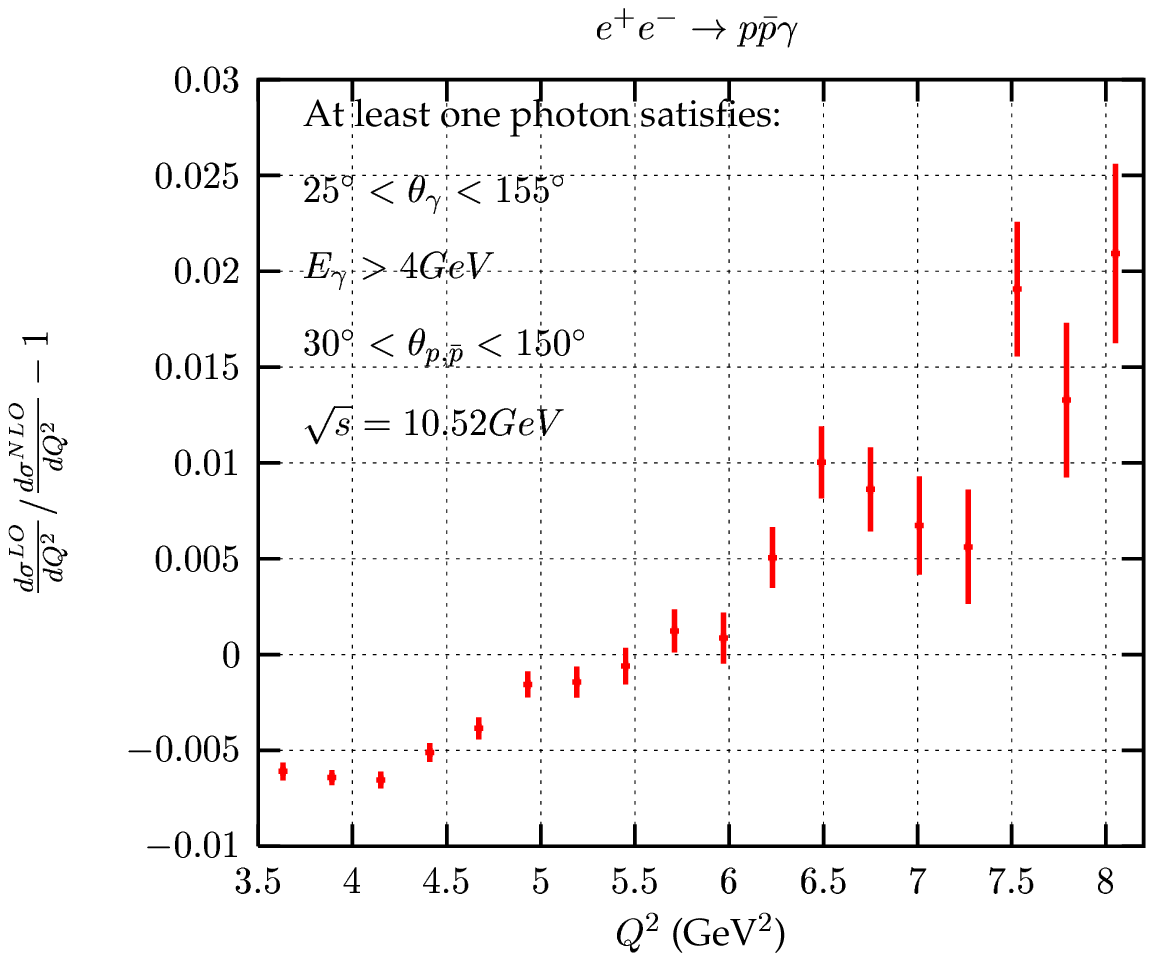,height=6.5cm,width=8cm} 
\epsfig{file=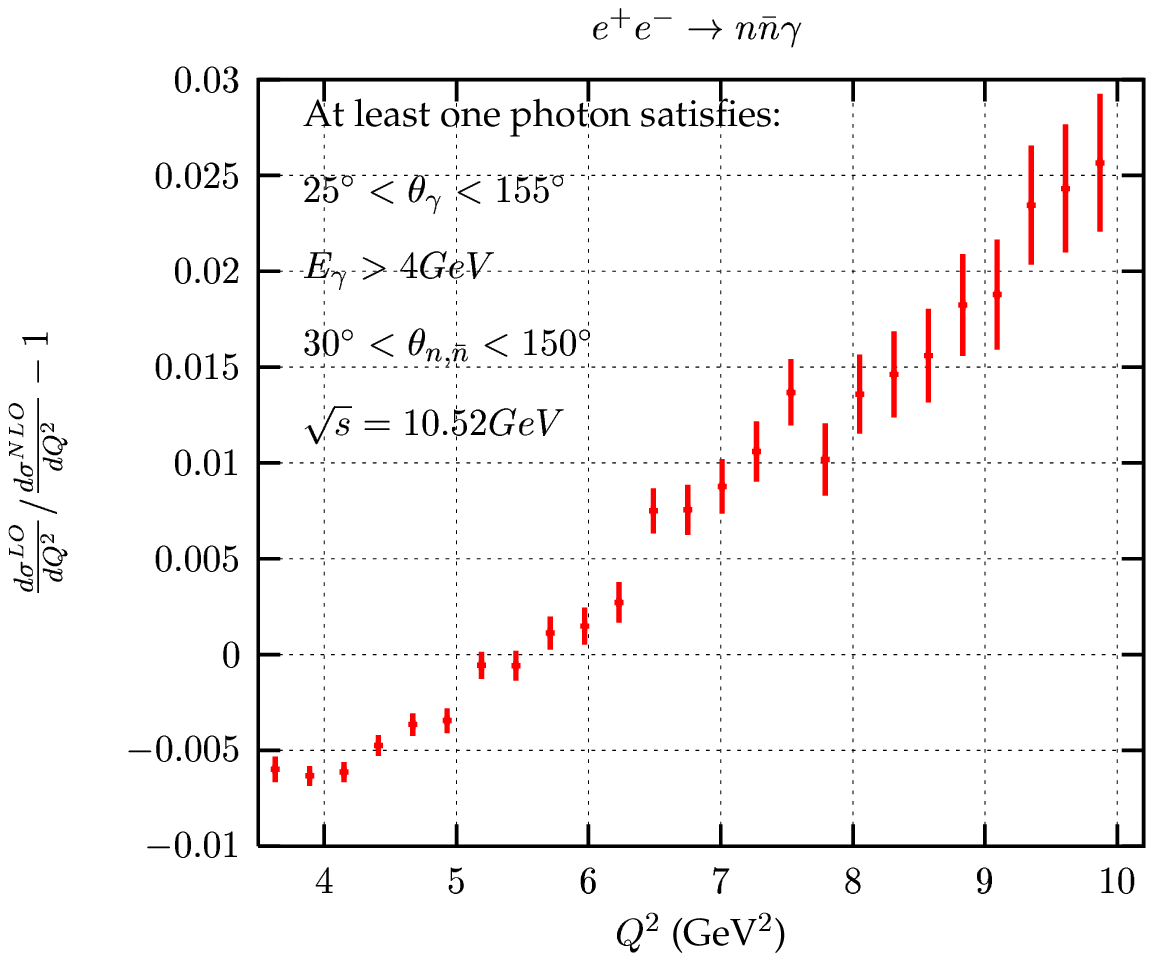,height=6.5cm,width=8cm}
\caption{Comparison between LO and NLO predictions for the 
differential cross section of the reactions
 $e^+e^-\rightarrow p \bar{p} \gamma$ 
 and $e^+e^-\rightarrow n \bar{n} \gamma$. }
\end{center}
\label{AA}
\end{figure}
\begin{figure}
\begin{center}
\epsfig{file=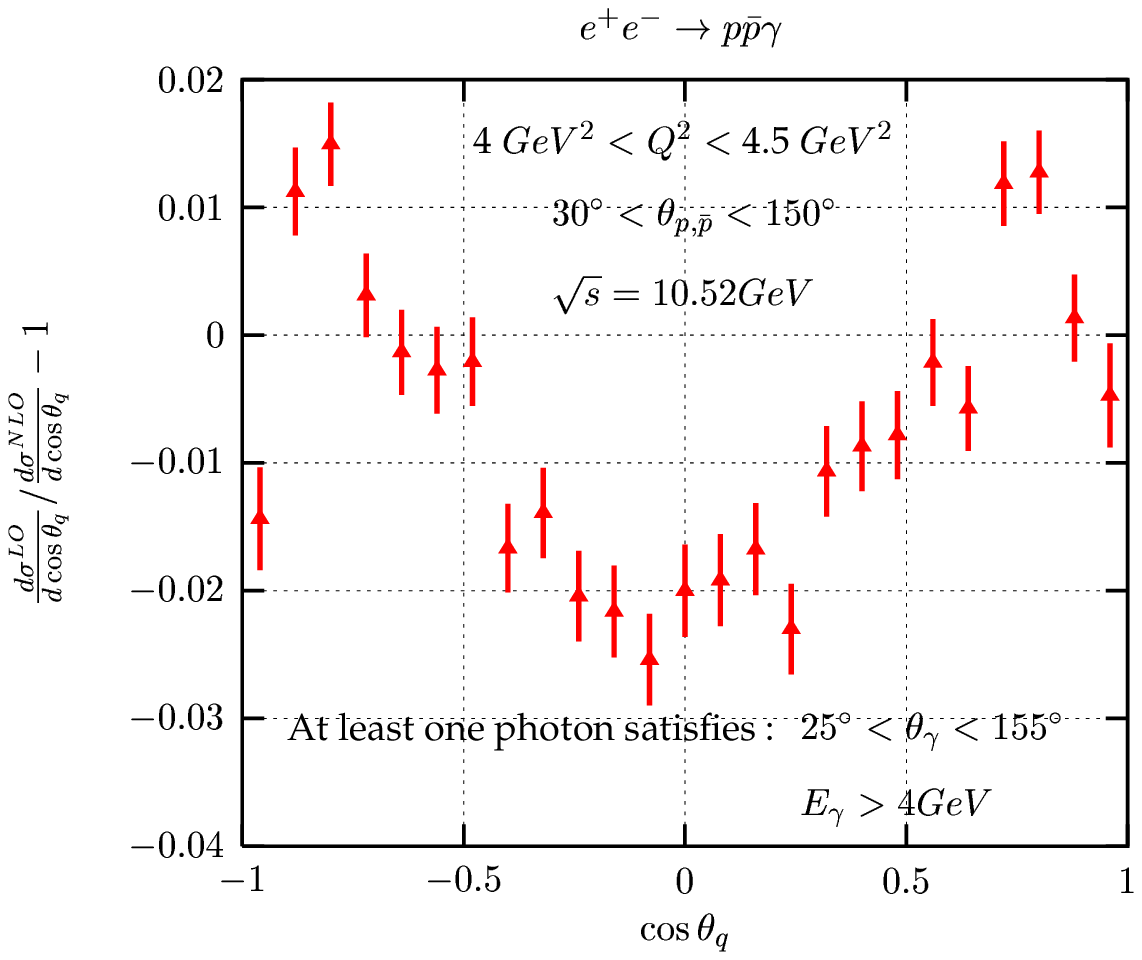,height=6.5cm,width=8cm} 
\epsfig{file=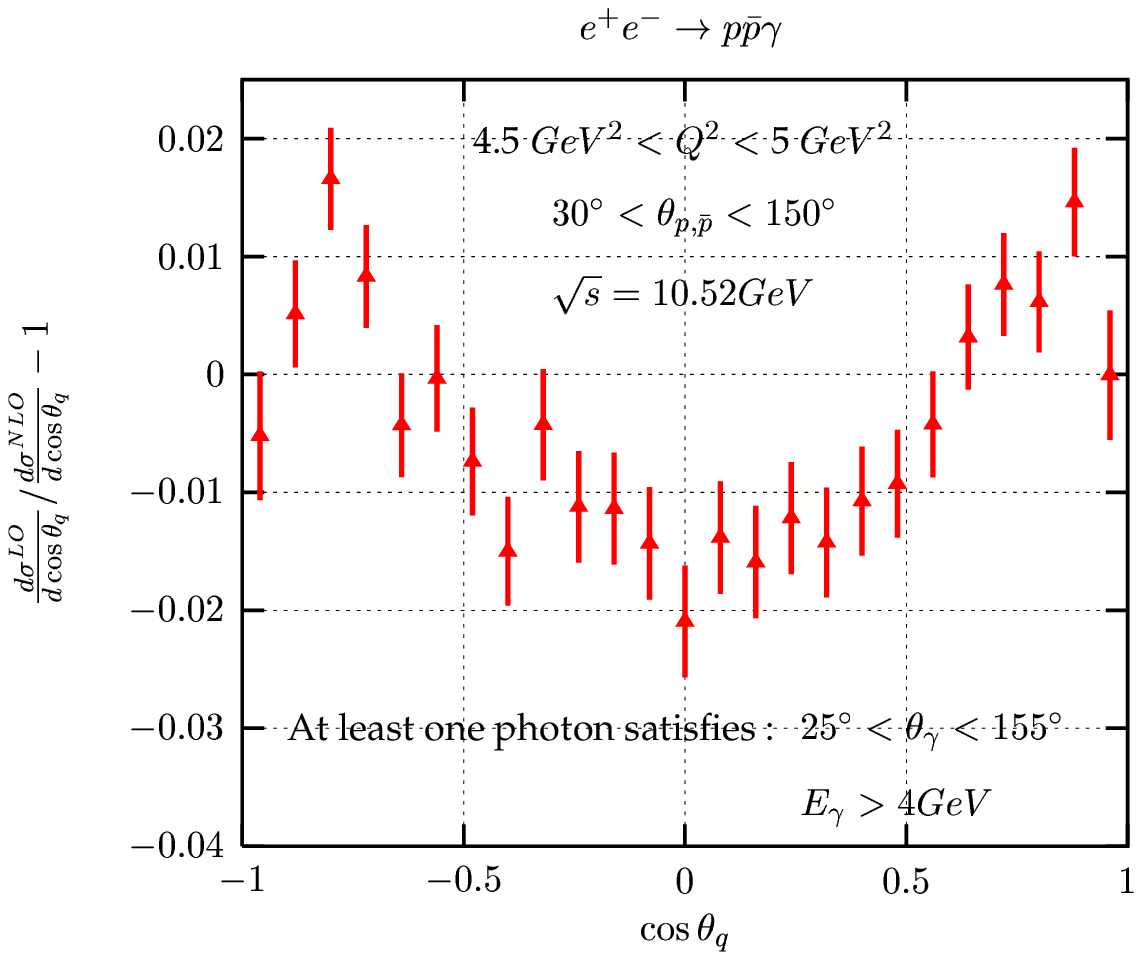,height=6.5cm,width=8cm}
\caption{Comparison between LO and NLO predictions for the 
differential cross section of the reaction
 $e^+e^-\rightarrow p \bar{p} \gamma$ for two different ranges of $Q^2$.}
\end{center}
\label{BB}
\end{figure}

The angular distribution  of the baryons, if defined  in the laboratory
or cms frame, is strongly  affected by the boost. The difference between
different choices for the  form-factor  ratios is  expected to  be more
pronounced  for the  proton angular  distribution in  the  hadronic rest
frame, with  the $z$-axis aligned with  the direction of  the photon and
the $y$-axis in the plane spanned  by the beam and the photon directions
(see Eqs.~(\ref{eq:LH2}) and (\ref{eq:LH3})  above). In this case  one may
study alternatively  the distributions with respect  to $\hat\theta$ and
$\hat\varphi$, the azimuthal and polar angles of the proton directions. For
$Q^2\ll s$, the case relevant to the present discussion, the $\hat\varphi$
dependence  is  unimportant,  and   the  pronounced  dependence  of  the
$\cos\hat\theta$  distribution on  the  choice of  the  form factors  is
clearly  visible.  The  difference between  the two  model  assumptions is
significantly more  pronounced, and the  sensitivity to the  form-factor
ratio improves (Fig. \ref{fig:15a}).

A quantitative estimate of the precision of such a cross section
measurement for proton production can be obtained from Fig. 6. As is
evident from Eq. (7), the relative error in the distribution $d\sigma/dQ^2$
for a specific $Q^2$-interval is a direct measure of the relative error
in $R(Q^2)$, averaged over the same interval. 
Taking, for example, the angular cuts for photons and protons
which roughly correspond to the detector acceptance, an integrated
luminosity of $100~{\rm fb}^{-1}$ and a bin size of $0.5~\rm{GeV}^{2}$,
one expects about 2500 / 600 / 250 events around 4 / 5 / 6 $\rm{GeV}^2$
and a resulting statistical precision of 2 / 4 / 6 \%. 
For $1000~{\rm fb}^{-1}$ even 8 ${\rm GeV}^2$ could be reached,
again with roughly 6\% statistical precision. For neutrons the rates are
even higher. 

The sensitivity to the ratio $|G_M/(\mu G_E)|$ can best be estimated from Fig.
9. Let us discuss the interval between 4.5 to 5 ${\rm GeV}^2$.
We expect in total around 1000 events, and the discrimination between
the two options shown in the Figure should be straightforward.

The importance of the radiative corrections can be deduced from Figs.\
10 and 11.  Even if the integrated cross sections are  similar in LO and
NLO (a difference below 0.5\% is observed in both Figs.\ 10 and 11), the
radiative  corrections do lead to distortions of the  distributions by
1--3\%; furthermore, they depend  on the  details of  the cuts  and the
criteria for  event selection. Hence, the  use of the  NLO generator is
highly recommended.

\section{Summary}

The radiative return at $B$-meson factories is well suited for 
measurements of the nucleon form factors over a wide kinematic range.
It should be emphasized that these measurements of the proton and
neutron form factors can be obtained from the data sample taken in
standard runs and close to the $\Upsilon$-resonance.
Valid for the time-like region, they would complement the precise results
for the space-like region from JLab. Close to the threshold the
statistical precision would be at the per cent level, and eventually,
depending on the integrated luminosity, these measurements could extend
out to 8 or even 9 ${\rm GeV}^2$.

In order to demonstrate the feasibility of the method for realistic
cuts, the Monte Carlo event generator PHO\-KHARA has been extended to simulate 
this reaction for $p\bar p \gamma$ and $n\bar n\gamma$ final states.
Examples for event rates and for angular distributions have been
presented, which include realistic cuts and which demonstrate the
feasibility of these measurements. NLO corrections to ISR amount to
typically 1--3\%. They are part of the present event generator
and should be included in a realistic simulation.

Angular distributions allow the separation of electric and magnetic
form factors. The general form of these distributions has been studied. 
They become relatively simple in the hadronic rest frame, with the 
$z$-axis aligned with the photon direction. For a specific (``optimal'')
orientation of the $z$-axis the leptonic tensor can even be diagonalized, 
which leads to a particularly simple form for the angular distribution.
In fact, this form is quite similar to the one observed in
the direct electron-positron annihilation reaction.
The simulation also demonstrates that the separation of
the squares of electric and magnetic form factors, respectively, can be
achieved over a fairly wide $Q^2$-range, even if realistic acceptance
cuts are imposed.  To get access to the relative phase between electric
and magnetic form factors, the determination of the nucleon spin is
required, which is an attractive option for $\Lambda \bar \Lambda$
production. We will come back to this possibility in a subsequent study.

\appendix

\section{The diagonal leptonic tensor} 

Let us start with the leptonic tensor (see e.g. \cite{Kuhn:2002xg}, Eq.~(4)) 
in the limit $m_e^2\ll s$ and $m_e^2/s \ll \theta^2_\gamma$: 
\begin{eqnarray}
L^{\mu\nu} &=& \frac{(4\pi\alpha)^2}{Q^4 y_1 y_2}
\biggl[-\left(2\frac{Q^2}{s} + y_1^2 + y_2^2\right) g^{\mu\nu} \nonumber \\
& &         - \frac{4 Q^2}{s^2}(p^\mu_1 p^\nu_1 + p^\mu_2 p^\nu_2)\biggr]
\label{eq:LEPT}
\end{eqnarray}
where $y_{1,2}=\frac{s-Q^2}{2s}(1\mp \cos\theta_{\gamma})$.

Terms proportional to $Q_\mu$ and $Q_\nu$ do not contribute as a
consequence of current conservation, $Q_\mu J^\mu_\mathrm{had}=0$, and have been
dropped. The momenta in the laboratory frame, with the $z$-axis pointing
along the positron beam are given by
\begin{equation}
p_{1,2}^\mu =\frac{\sqrt{s}}{2} 
        \left(\begin{array}{c} 1\\0\\0\\ \pm 1 \end{array} \right), 
\;\;\;
k^\mu =\frac{s-Q^2}{2\sqrt{s}} 
        \left(\begin{array}{c} 1\\0\\s_\gamma\\ c_\gamma \end{array} \right),
\nonumber
\end{equation}
\begin{equation}
Q^\mu =\frac{s-Q^2}{2\sqrt{s}} 
        \left(\begin{array}{c} \frac{s+Q^2}{s-Q^2}\\0\\
       - s_\gamma\\ -c_\gamma \end{array} \right), 
\label{momdef}
\end{equation}
with $s_\gamma = \sin\theta_\gamma$ and $c_\gamma= \cos\theta_\gamma$ 
accordingly. We perform the following transformations: a rotation of
the coordinate frame around the $x$-axis, such that the new $z$-axis
points into the negative photon direction and subsequent boost from the
laboratory frame along the new $z$-axis with $\gamma\equiv
(s+Q^2)/2\sqrt{s Q^2}$ into the hadron rest frame. Then, one finds
\begin{equation}
p_{1,2}^\mu =\frac{\sqrt{s}}{2}\gamma 
        \left(\begin{array}{c} (1\pm\beta c_\gamma)\\
            0\\ \pm s_\gamma/\gamma\\ (-\beta\mp c_\gamma)
              \end{array} \right), 
\;\;\;\;
q^\mu =\frac{\sqrt{Q^2}}{2}\beta_N 
        \left(\begin{array}{c} 0\\
                 s_{\hat\theta}c_{\hat\varphi}\\
                 s_{\hat\theta}s_{\hat\varphi}\\              
                 c_{\hat\theta}
         \end{array} \right),
\nonumber
\end{equation}
\begin{equation}
Q^\mu =\sqrt{Q^2} \left(\begin{array}{c} 1\\0\\0\\0\end{array} \right).
\label{momlabdef}
\end{equation}
In this coordinate system the space components of the leptonic tensor are 
given by
\begin{eqnarray}
L^{ij} &=& \frac{(4\pi\alpha)^2}{Q^4 y_1 y_2}
\biggl[\left(2\frac{Q^2}{s} + y_1^2 + y_2^2\right) \delta^{ij} \nonumber \\
& &           - \frac{Q^2}{s}\gamma^2(v^i_1 v^j_1 + v^i_2 v^j_2)\biggr]~,
\label{eq:LEPThad}
\end{eqnarray}
with
\begin{equation}
 \vec v_{1,2} =\left(\begin{array}{c} 0\\
             \pm s_\gamma/\gamma\\ (-\beta\mp c_\gamma)
              \end{array}\right). 
\end{equation}

In combination with the had\-ronic
tensor this leads immediately to Eq.~(\ref{eq:LH2}).
The leptonic tensor, as given above, is evidently symmetric.
The combination
\begin{equation}
(v^i_1 v^j_1 + v^i_2 v^j_2)/2=\left(\begin{array}{lll}
0 \phantom{000}& 0                         &        0 \\
0 & s^2_\gamma/\gamma^2        & -s_\gamma c_\gamma/\gamma \\
0 & -s_\gamma c_\gamma/\gamma\phantom{0}  & (\beta^2 + c_\gamma^2) \\
\end{array} \right )
\end{equation}
can thus be diagonalized by a rotation of the coordinate system
around the $x$-axis, now, however, in the hadron rest frame. 
The rotation angle $\theta_D$ is obtained from
\begin{equation}
\tan \theta_D = \sqrt{\frac{\lambda - a}{\lambda + a}}
   = \frac{\gamma(\lambda - a)}{2s_\gamma c_\gamma },
\end{equation}
or, alternatively
\begin{equation}
\tan(2\theta_D) = \frac{2s_\gamma c_\gamma }{\gamma a}
\end{equation}
where
\begin{equation}
\lambda = \lambda(\beta^2,-c^2_\gamma, s^2_\gamma/\gamma^2);\;\;\;\;
a=(\beta^2 + c^2_\gamma - s^2_\gamma/\gamma^2),
\end{equation}
with
\begin{equation}
\lambda(x,y,z)=\sqrt{x^2+y^2+z^2 - 2 (xy + xz + yz) }.
\end{equation} 
The eigenvalues of $(v^i_1 v^j_1 + v^i_2 v^j_2)/2$ are given by
\begin{equation}
\lambda_\pm = (\beta^2 + c^2_\gamma + s^2_\gamma/\gamma^2 \pm\lambda)/2 
\end{equation}
and the leptonic tensor in the new ``optimal'' frame simplifies to
\begin{eqnarray}
L^{ij} &=& \frac{(4\pi\alpha)^2}{Q^4 y_1 y_2}
\biggl[\left(2\frac{Q^2}{s} + y_1^2 + y_2^2\right) \delta^{ij} \nonumber \\
& &   - 2\frac{Q^2}{s}\gamma^2 \mbox{\rm diag}(0,\lambda_-,\lambda_+)\biggr]~.
\end{eqnarray}

Choosing this new ``optimal''
frame for the definition of the baryon direction with
angles denoted by $\tilde\theta$ and $\tilde\varphi$, only quadratic
terms in $\cos\tilde\theta$ and $\sin\tilde\theta\sin\tilde\varphi$
remain in the angular distribution.
The combination $L_{ij}H^{ij}$ is now given by
\bea
&&L^{ij}H_{ij}= \frac{(4\pi\alpha)^2}{Q^2 y_1 y_2}
\biggl[ (2\frac{Q^2}{s} + y_1^2 + y_2^2)(3A + \beta_N^2 B) 
-2 \frac{Q^2}{s} \gamma^2 \nonumber\\ 
&& \times 
 \biggl( A (\lambda_+ + \lambda_-) +
\beta^2_N B 
(\lambda_- \sin^2\tilde\theta\sin^2\tilde\varphi
+\lambda_+\cos^2\tilde\theta)\biggr) \biggr]~, \nonumber \\
\eea
where $A$ and $B$ characterize the hadronic tensor
\bea
H_{\mu\nu}&=&A (Q_\mu Q_\nu - g_{\mu\nu} Q^2) 
+4 B q_\mu q_\nu \ .
\label{had_tens_gen}
\eea
For two-particle production this is the most general form
of the symmetric hadronic tensor,
if one sums over polarizations of the particles in the final state.
For the nucleon--antinucleon final state specifically, one finds
\bea
 \kern-10pt
 A =  2|G_M^N|^2 \ \ {\rm and}
 \ \  B = - \frac{2\tau}{\tau -1}\left(|G_M^N|^2-\frac{1}{\tau}|G_E^N|^2\right).
\label{AB}
\eea
Let us emphasize that this choice of coordinates also leads to an
extremely simple angular distribution for pion pairs, 
and quite generally for any single-particle inclusive distribution.

In the the limit of $Q^2\ll s$ the rotation angle $\theta_D$ vanishes,
the direction of the photon can be used as (negative) $z$-axis
and the leptonic tensor reduces to
\begin{equation}
L^{ij}=\frac{(4\pi\alpha)^2}{Q^4 y_1 y_2}
\frac{(1+c_\gamma^2)}{2} \mbox{\rm diag}(1,1,0)~,
\end{equation}
and
\begin{equation}
L^{ij}H_{ij} \simeq \frac{(4\pi\alpha)^2}{Q^2 y_1 y_2}
 \left(1+c_\gamma^2\right) \left[2A +B\beta^2_N\sin^2\tilde\theta\right]\ ,
\end{equation}
which for $A$ and $B$ from Eq.~(\ref{AB}) gives
\bea
&& \kern-30pt L^{ij}H_{ij} \simeq \frac{(4\pi\alpha)^2}{Q^2 y_1 y_2}
 \left(1+c_\gamma^2\right) \nonumber \\
&&\times \left[|G_M^N|^2\left(1+\cos^2\tilde\theta\right)
 +\frac{1}{\tau}|G_E^N|^2\sin^2\tilde\theta \right] \ 
\eea
which is closely related to Eq.~(\ref{sigdir}).



\end{document}